\documentclass[conference]{IEEEtran}
\usepackage[hyphens]{url}
\usepackage{hyperref}
\hypersetup{colorlinks,allcolors=blue}

\usepackage[backend=bibtex,style=ieee,citestyle=numeric-comp]{biblatex}
\IEEEoverridecommandlockouts

\usepackage{graphicx} 
\usepackage{xcolor} 
\usepackage{minted} 
\usepackage{doi}
\def\BibTeX{{\rm B\kern-.05em{\sc i\kern-.025em b}\kern-.08em
    T\kern-.1667em\lower.7ex\hbox{E}\kern-.125emX}}

\begin{document}

\title{Bridging the Smart City Cybersecurity Data Gap Through AI-Driven Synthetic Dataset Generation}

\author{\IEEEauthorblockN{Stephanie Polczynski, John D. Hastings, Varghese Vaidyan, Kyle Korman}
\IEEEauthorblockA{\textit{The Beacom College of Computer \& Cyber Sciences} \\
\textit{Dakota State University}\\
Madison, SD, USA \\
stephanie.polczynski@trojans.dsu.edu, john.hastings@dsu.edu, varghese.vaidyan@dsu.edu, kyle.korman@dsu.edu
}
}

\maketitle

\begin{abstract} 
Smart cities rely on interconnected cyber-physical systems that integrate sensors, IoT devices, cloud platforms, and AI-driven services and decision-making. While these systems enhance city services, they also introduce complex cybersecurity challenges due to their large attack surfaces, heterogeneous data flows, and evolving threat vectors. Developing and validating cybersecurity tools for smart cities requires high-quality datasets that accurately represent real operational conditions. However, real-world datasets are often incomplete, contain privacy-sensitive data, are difficult to access, or lack sufficient malicious activity to support tool development. This research addresses this critical gap by proposing an AI-based synthetic data generation (SDG) framework designed specifically for smart city cybersecurity research. The proposed framework leverages generative artificial intelligence models to produce high-fidelity synthetic cybersecurity datasets that replicate realistic device behaviors, network interactions, and cyber-attack scenarios. The synthetic datasets are evaluated for conformity to protocol standards, statistical similarity to original datasets, and utility in common security tools. The resulting synthetic data generation framework and evaluation metrics are expected to advance smart city cybersecurity by enabling researchers to model threats more effectively and evaluate defensive techniques more comprehensively to better protect critical smart city infrastructures.  
\end{abstract}


\section{Introduction}
Smart cities are a rapidly growing technology area where the combination of information and communication technologies, edge computing, AI/ML, and other disciplines are combined to digitize and enhance city services and safety for a community. The blending of several complex technologies, the cyber-physical nature of some of the services, and the vast amounts of data collected and processed to enable smart city services create unique cybersecurity and privacy challenges.

Smart city technologies had over a nine billion dollar market size in the US in 2024 \cite{grandviewresearch_2025}. However, even with the rapid growth of the field and the increasing buzz for the convergence of technologies that enable smart cities, it can be difficult to pinpoint what exactly it means to have a smart city. ITU defines a smart city as “an innovative city that uses information and communication technologies (ICTs) and other means to improve quality of life, efficiency of urban operation and services, and competitiveness, while ensuring that it meets the needs of present and future generations with respect to economic, social, environmental as well as cultural aspects” \cite{itu_2022}. Others define smart cities as installing digital interfaces in traditional city infrastructure or streamlining city operations by using technology and the resulting data to optimize city services and provide a better quality of life for its citizens \cite{woetzel_2018}.

While it can be difficult to pinpoint what constitutes a smart city, it is clear that smart cities rely on the intersection of technology, government, and human aspects for solutions to be successful. Technologically, a smart city requires integration of sensors, networking, and data analytics for successful operation. These are further powered by a complex ecosystem of telecommunications infrastructure, IT networks, existing city infrastructure systems, and automation and building control. \cite{lea_2017} highlights the following areas as enablers of smart cities: networking and communication technologies such as Low-Power Wide Area Networks (LPWAN) and cellular Internet of Things (IoT) networks, the increasing prevalence of cyber-physical and IoT systems for improved sensing and actuation of city services, cloud and edge computing to handle the large amounts of data and required compute cycles, open data policies to allow third-parties to analyze the city to enable the improvement and creation of new city services, data analytics to offer insights that can improve efficiency, and finally citizen engagement because smart cities do not work without the citizens participating and taking advantage of the improved city services .

When combining the driving technologies of smart cities, all of which have cybersecurity issues of their own, and overlaying them with essential city services such as transportation, power, water, healthcare, and public safety, there are bound to be complex cybersecurity issues to address. \cite{houichi} provided four major cybersecurity challenges for smart cities: sophisticated attacks, software product bugs, vulnerabilities, and legislative issues. Cybersecurity threats in the form of data breaches, insider threats from smart city government maintainers or the vendors running the technologies, denial-of-service attacks, abuse of open data, and other malicious attacks may deny, disrupt, or degrade essential city services. There are no shortage of cyber threats to research and mitigate when it comes to smart cities, though getting started in the area can be challenging due to the complex interconnected nature of smart cities, the large amounts of data processed, and the difficulty in getting access to real-world systems and data.

In preliminary research, a survey on smart city cybersecurity and privacy current research was conducted with a focus on exploring known vulnerabilities towards smart city technologies, examining smart city resilience, identifying existing guides and frameworks aimed towards ensuring the cybersecurity of smart cities, and examining data privacy issues introduced by the massive amounts of data collected and processed by smart cities. A conclusion of this research was that the current state of research in smart city cybersecurity problems can be characterized as of poor research quality partially due to the lack of available real world data and testing tools.

Further preliminary research was also conducted into examining the current state of existing smart city datasets. Existing academic research in this area focuses on describing available datasets and the types of data available in them. A new approach was taken to quantify characteristics of the datasets, to include metrics such as: size, formats, data types, domains, update frequency, etc. While this proved valuable to understanding the available datasets and their potential uses, a notable finding was that most publicly available smart city datasets differ from those commonly used in cybersecurity research, which often include processed and curated data focused on specific security events or incidents. Instead, the smart city datasets examined were more operational in nature, encompassing a wide range of city metrics and are often composed entirely of raw data and measurements. It would likely prove challenging to perform effective cybersecurity research using this data.

\section{Problem Statement}
Based on the above findings, the problem that this research will address is that there is both a lack of real world data available for cybersecurity researchers and that the existing datasets do not contain the type of data typically present in cybersecurity datasets. Deployers and maintainers of smart city technologies are unlikely to be willing to share real-world datasets due to wanting to protect the security and privacy of their infrastructure. Creating a technique for generating representative data that can be freely shared will enable future researchers to advance the field of smart city cybersecurity.

This research is designed to close the gap in smart city cybersecurity dataset availability through generating synthetic data for smart cities to create training data that could be used by future cybersecurity researchers or for testing security tools for smart cities. To achieve this goal, it will also be necessary to gain a fundamental understanding of what constitutes a smart city cybersecurity dataset, examine existing approaches for generating synthetic data, develop evaluation techniques for determining successful synthetic dataset generation, and to create a means to introduce known bad activity into the dataset.

\section{Research Questions}

These goals will be accomplished through the following research questions:

\begin{noindent}
\textbf{RQ1: What properties of a smart city cybersecurity dataset are the most meaningful for future research and tool development?}
\end{noindent}

A preliminary research task will be defining what constitutes such a dataset. What datatype, features, and other characteristics should be present in such a dataset for it to be meaningful for use?

Work will also need to be done to determine the minimum characteristics of smart city technologies that need to be included in synthetic data to ensure accurate representation of real-world data sets.

\begin{noindent}
\textbf{RQ2: Can generative AI be used to create representative synthetic smart city datasets? } 
\end{noindent}

This research question will also require two lines of effort. The first is to explore existing AI/ML techniques for generating representative data and evaluate each for feasibility in the smart city use case. Existing synthetic data generation (SDG) methods will be used to create sample synthetic smart city cybersecurity datasets. 

Once the synthetic datasets have been generated, there will also need to be an analysis to ensure that they accurately represent real-world data and that they will be meaningful for cybersecurity researchers. This research will develop methodology for evaluating the reliability and accuracy of the synthetic datasets.

\begin{noindent}
\textbf{RQ3: Can realistic attack scenarios be accurately created through synthetic data generation? }
\end{noindent}

Finally, with known synthetic datasets in hand that have been determined to be representative and accurate compared to real-world data, the final research question is can malicious behavior and/or attack scenarios be introduced into the SDG process. This is needed because without known bad data, security tools cannot be reliably trained or trusted. 
\section{Related Work}
This section surveys the current state of smart city data resources and the emerging role of SDG in addressing persistent gaps in availability, realism, and security relevance. It begins by examining the characteristics, limitations, and challenges of existing smart city datasets. The section will then transition to SDG techniques and applications for technologies that are present in or adjacent to smart city networks. A review of existing techniques for assessing the fidelity, utility, and protocol correctness of synthetic data will be presented. The section will conclude by discussing gaps in both smart city data resources and SDG methodologies.

\subsection{Smart City Cybersecurity Datasets}

A major challenge in working with smart city data is ensuring data quality while managing its diverse characteristics. As \cite{Mallapuram_Ngwum_Yuan_Lu_Yu_2017} note, sensor data often appears in many formats, requiring systems capable of handling large-scale, heterogeneous inputs. Collecting the data itself can also be difficult, as sensors are frequently resource constrained and network infrastructure may be insufficient to transmit high-volume data to centralized repositories. \cite{Ma_Preum_Ahmed_Tärneberg_Hendawi_Stankovic_2020} further highlight challenges such as heterogeneity in sampling rates and aggregation levels—for example, real-time traffic data versus monthly pollution measurements. Smart city deployments also vary in focus, making it difficult to generalize patterns or design platforms that serve multiple cities. Data formats differ widely across sensors (e.g., audio, video, text, raw telemetry), and the interdisciplinary nature of smart city applications requires expertise across domains such as computer science, statistics, transportation, and urban planning. Ensuring data integrity, completeness, and timeliness remains difficult due to the distributed nature of sensors and the complexity of maintaining large-scale, geographically dispersed systems.

\subsubsection{Existing Datasets}

Smart city datasets in this research fall into two categories: operational datasets and cybersecurity datasets. Operational datasets capture measurements from city services—such as energy use, traffic flow, environmental conditions, waste management, and citizen engagement—and are primarily used to improve service efficiency and urban planning. Cybersecurity datasets, by contrast, contain network traffic logs, intrusion detection outputs, and real or simulated attack traces that support the detection and analysis of malicious activity targeting smart city infrastructure.

There are several existing papers documenting past and present smart city datasets. Perhaps the most extensive approach at this is from \cite{Ma_Preum_Ahmed_Tärneberg_Hendawi_Stankovic_2020}, with 14 smart cities listed and available datasets provided within the domains of transportation, emergency \& public safety, energy, environment, and social sensing. Upon manually reviewing each of these real-world datasets presented, the existing formats all appear to contain operational data only and are not data relevant to cybersecurity research.

When looking specifically at cybersecurity datasets, there are several lab-generated options to use as a starting point. In the past few years, there has been substantial efforts to create datasets for use in IoT and Industrial Internet of Things (IIoT) research. While these are not directly targeted towards smart city applications, the datasets contain devices, protocols, technologies, and deployment methods that are typically present in individual systems running smart city services. For example, the Edge-IIoTSet  \cite{Ferrag_Friha_Hamouda_Maglaras_Janicke_2022} contains IIoT devices and protocols from industrial systems. This may be useful when defining what a smart city cybersecurity dataset should contain and for building representative traffic for individual components of a smart city. Further, some of these datasets also include known malicious behavior, which may prove useful for generating known bad behavior in the synthetic datasets. A summary of available lab-created IoT/IIoT cybersecurity datasets in provided in Table \ref{tab:iot_iiot_datasets}.

\begin{table*}[hbtp!]
\centering
\resizebox{\textwidth}{!}{
\begin{tabular}{|p{2.5cm}|p{1cm}|p{3.5cm}|p{3.5cm}|p{3.5cm}|p{3.5cm}|}
\hline
\textbf{Dataset}&  \textbf{Year}&  \textbf{Purpose}&  \textbf{Generation}&  \textbf{Content}& \textbf{Key Features}\\\hline
Edge-IIoTSet \cite{Ferrag_Friha_Hamouda_Maglaras_Janicke_2022}& 2022& Designed to provide a benchmark for IDS systems in IoT and IIoT applications& Created from a seven-layer testbed that included multiple IoT devices, IIoT specific protocols (e.g., Modbus), and cloud and edge deployments& 72M+ records in PCAP and flow formats. Includes 14 types of attack scenarios&Structured to support centralized and federated learning approaches\\\hline
 WUSTL-IIoT-2021\cite{Zolanvari_Teixeira_Gupta_Khan_Jain}& 2021& Designed specifically for cybersecurity research in IIoT& Collected from an IIoT tesbed at Washington University that emulates real-world ICS systems& \~1.2M records with 41 features, captured over 53 hours&Includes attacks relevant to industrial environments, such as DoS, command injection, and reconnaissance\\\hline
 X-IIoTID\cite{Al-Hawawreh_Sitnikova_Aboutorab_2022}& 2022& Connectivity- and device-agnostic dataset for IIoT intrusion detection& Created from an three-tiered IIoT system with industrial devices, edge computing, mobile devices, and cloud services& Contains standard network traffic and IIoT protocols such as MQTT, CoAP, and WebSocket&A simulation of recent attacker TTPs was used to generate malicious behavior. Contains multi-view data from network traffic, system and application logs, host metrics, and IDS logs.\\\hline
UNSW\_NB15 \cite{Moustafa_Slay_2015}& 2015& Benchmark IDS, network analysis, and ML models on realistic, labeled network traffic& Traffic generated using IXIA PerfectStorm in test environment, combining benign and attack traffic& PCAP files plus extracted flow features with labels& Multiple attack categories and detailed feature sets \\\hline
 BoT\_IoT \cite{Koroniotis_Moustafa_Sitnikova_Slay_2018}& 2020& Benchmarking and training cybersecurity systems against IoT attack scenarios& Collected from IoT testbed running smart devices under normal and attack scenarios& 72M+ records. PCAP captures, processed flow features in CSV, attack and benign traffic labels& High volume, multiple attack types, with clear labeling \\\hline
 TON\_IoT \cite{Moustafa_Keshky_Debiez_Janicke_2020}& 2021& Evaluation of AI/ML-based cybersecurity tasks in IoT/IIoT contexts& Collected from a dedicated cyber range combining IoT, fog/edge/cloud layers, multiple OS environments, and organized attacks& Network traffic, telemetry logs, OS audit traces, and processed features with labels& Heterogeneous sources, multiple attack categories, train/test splits, and ground truth labeling \\\hline
    \end{tabular}
    }
    \caption{Existing IoT/IIoT Cybersecurity Datasets}
    \label{tab:iot_iiot_datasets}
\end{table*}

\subsection{Synthetic Data Generation}

Synthetic data is "artificially generated data that mimics the characteristics and patterns of real-world data, but is created through algorithms, generative models, or simulations" instead of being directly created by humans or real-world systems \parencite{Liu_Wei_Liu_Si_Zhang_Rao_Zheng_Peng_Yang_Zhou_etal._2024}. Within cybersecurity, synthetic data provides a controlled means to model normal activity, emulate attacks, and test defensive tools at scale, making it a growing capability for advancing intrusion detection, anomaly detection, and other cybersecurity research fields when high-quality, labeled datasets are unavailable. The creation and use of synthetic data allows researchers to overcome limitations associated with scarce, proprietary, or privacy-restricted datasets, all of which are challenges that are especially prominent in smart city and IoT networks. 

\subsubsection{Current Methods}

\cite{Osorio-Marulanda_Epelde_Hernandez_Isasa_Reyes_Iraola_2024} conducted a systematic literature review of SDG techniques, categorizing them into four primary groups: GAN-based, machine learning-based, statistical-based, and kernel-based approaches. Similarly, \cite{Viana_Teixeira_Baptista_Pinto_2024} compiled a comprehensive overview of SDG methods, with a particular emphasis on statistical techniques. Collectively, these works provide a structured foundation for understanding the major approaches used in SDG. Among these, Generative Adversarial Networks (GANs) have emerged as the most prominent, leveraging adversarial training between a generator and discriminator to produce realistic synthetic data. Since their introduction in 2014, GANs have evolved into numerous variants to address challenges such as training stability, privacy, and domain specificity. Examples include WGAN for improved training stability, DPGAN for incorporating differential privacy, and DCGAN for hierarchical feature learning using convolutional architectures. Their flexibility and ability to model complex, high-dimensional data distributions have made GANs widely adopted across domains.

In contrast, machine learning-based techniques rely on traditional algorithms such as decision trees, support vector machines, and ensemble methods to learn patterns and generate synthetic data, often prioritizing interpretability and control over the generation process. Statistical methods, including Gaussian mixture models, Bayesian networks, and copulas, generate synthetic data by sampling from estimated probability distributions, preserving key statistical properties such as means, variances, and correlations while offering simplicity and strong theoretical grounding. Finally, kernel-based approaches utilize similarity functions, such as kernel density estimation, to capture non-linear relationships and local data structures, enabling the generation of synthetic samples that closely resemble the original data distribution. While less commonly used than GANs or statistical methods, kernel-based approaches provide fine-grained control and are particularly valuable in scenarios requiring nuanced handling of variability and privacy. Together, these SDG techniques represent a diverse set of methodologies, each with distinct strengths and trade-offs depending on the application domain and data characteristics.

\subsubsection{Current Applications of SDG for Cybersecurity}

Recent research demonstrates the growing application of SDG to address cybersecurity challenges across various domains. For example, \cite{Wang_Govindarasu_2024} proposed a GAN-inspired framework integrated with the Fast Gradient Sign Method (FGSM) to generate adversarial synthetic datasets for anomaly detection in Distributed Energy Resource (DER) networks using the DNP3 protocol. This approach effectively mitigates data imbalance issues and improves classification performance in supervised learning models, outperforming traditional techniques such as SMOTE and random undersampling. Similarly, \cite{Singh_Bettouche_Fischer_2024} explored SDG for addressing data scarcity in process mining, evaluating both LSTM and GAN-based models. Their findings highlight a trade-off between methods, where LSTMs provide higher fidelity data reproduction, while GANs enhance robustness and variability, demonstrating the value of SDG in improving machine learning performance under limited data conditions.

Other studies focus on expanding SDG capabilities to handle complex and heterogeneous data environments. \cite{Tenison_Chen_Singh_Dahleh_Zemour_Kagal_2024} introduced a modular pipeline for generating mixed-type datasets, combining GAN-based models for structured data with privacy-aware large language models (LLMs) for unstructured text, enhanced through Named Entity Recognition (NER) to improve contextual fidelity. This work emphasizes the importance of privacy-preserving SDG in real-world applications. Additionally, earlier work by \cite{Myung_Choi_Choi_Yu_Lee_Lee_2016} demonstrated a statistical approach to SDG in smart home environments, using techniques such as the Apriori algorithm and Markov chains to model user behavior and generate realistic service interaction data. Together, these applications illustrate the breadth of SDG use in cybersecurity, spanning adversarial data generation, privacy-preserving pipelines, and behavior modeling, while highlighting the adaptability of different techniques to domain-specific challenges.

\subsubsection{Synthetic Data Generation for Network Traffic}

Recent advancements in SDG for network traffic focus on improving realism by capturing temporal dynamics, protocol behavior, and interactions between network entities. \cite{Wu_Chen_Chou_Wang_2026} introduced SPATGAN, a multi-agent GAN framework that models bidirectional client–server interactions using separate generators for timing and packet features, enabling more accurate representation of request–response behavior. Similarly, IoTGemini \cite{Li_Li_Zou_Zhao_Zeng_Huang_Jiang_Lyu_Ormazabal_Singh_et} employs a two-stage approach combining device-level behavioral modeling with a packet-sequence GAN to preserve both per-packet attributes and sequential dependencies. These approaches demonstrate significant improvements in distributional similarity and downstream task performance, such as intrusion detection and anomaly detection, highlighting the importance of modeling both structural and temporal characteristics of network traffic.

Other works explore alternative architectures and domain-specific adaptations for synthetic traffic generation. \cite{Jiang_Wang_Liu_Xu_Zou_Zhang_Tan_Zhang_2025} proposed a conditional GAN framework for generating network-flow data in non-terrestrial IoT environments, integrating embedding techniques to capture relationships between flow attributes and application behaviors. In contrast, \cite{Schoen_Blanc_Gimenez_Han_Majorczyk_Me_2024} demonstrated that Bayesian Networks can outperform GAN-based approaches in preserving protocol semantics, feature dependencies, and overall data realism, emphasizing the importance of model selection based on data characteristics and use cases.

Finally, SDG has been extended to support cybersecurity-specific objectives such as anomaly detection and attack generation. \cite{Fioretto_Masciari_Napolitano_2025} developed a synthetic WiFi traffic generator integrated with an explainable anomaly detection pipeline, enabling privacy-preserving experimentation with controllable anomalies. Likewise, \cite{Nguyen_Le_Le-Minh_Le_2023} introduced DGIDS, a semi-synthetic framework that combines real benign traffic with GAN-generated attack data to produce high-quality labeled datasets. Their results show substantial improvements in intrusion detection performance, demonstrating the value of incorporating realistic attack scenarios into synthetic data. Collectively, these approaches highlight the evolution of SDG toward more realistic, scalable, and application-driven network traffic generation for cybersecurity research.

\subsection{Evaluating Synthetic Data}

\subsubsection{Accuracy and Representativeness}

Evaluating synthetic data for accuracy and representativeness is essential to ensure that generated datasets faithfully reflect the statistical properties, relationships, and practical utility of the original data. This evaluation typically examines how well synthetic data preserves distributions, correlations, and task-specific performance while balancing privacy risk. Table \ref{tab:synthetic_data_evaluation_tools_comparison} summarizes widely used open-source synthetic data evaluation frameworks and tools, highlighting their primary focus areas, evaluation dimensions, strengths, and limitations.

\begin{table*}[htbp!]
\centering
\resizebox{\textwidth}{!}{
\begin{tabular}{|p{3.0cm}|p{4.2cm}|p{4.2cm}|p{4.2cm}|p{4.2cm}|}
\hline
\textbf{Tool} & 
\textbf{Primary Focus} &
\textbf{Evaluation Dimensions} &
\textbf{Strengths} &
\textbf{Limitations} \\\hline

\textbf{SDNist} \cite{Task_Bhagat_Howarth_2023} &
Standardized synthetic data evaluation &
Utility, privacy risk, statistical similarity &
Highly recognized; strong privacy risk guidance; structured and repeatable &
Requires manual documentation; limited automation  \\\hline

\textbf{SDMetrics} \cite{SDMetrics_2025} &
Automated quality evaluation for tabular synthetic data &
Fidelity, utility, correlation, distribution similarity, privacy metrics &
Highly automated; extensive metric library; easy Python integration &
Best for tabular data; limited domain-specific metrics  \\\hline

\textbf{SynthEval} \cite{schneiderkamplab/syntheval_2025} &
Evaluate the quality of synthetic data compared with real data &
Utility, privacy, classification, fairness &
Allows creation of new metrics & 
Limited support for sequential or time-series data \\\hline

\textbf{SynthRO} \cite{Santangelo_Nicora_Bellazzi_Dagliati_2025,bmi-labmedinfo/SynthRO_2025} &
Visual dashboards for benchmarking metrics &
Similarity, privacy, utility &
Modular structure allows for new evaluation metrics to be added &
Tailored to healthcare and privacy use cases \\\hline
\end{tabular}
}
\caption{Comparison of Open-Source Synthetic Data Evaluation Tools}
\label{tab:synthetic_data_evaluation_tools_comparison}
\end{table*}

\subsubsection{Benchmarks for Determining Success}

Metrics for assessing synthetic data accuracy increasingly focus on whether generated samples preserve the statistical structure, temporal behavior, and relational patterns of the real data they aim to replace. Traditional measures such as column‑wise similarity scores and correlation preservation capture how well individual features or distributions align, but they often miss the joint dynamics that matter in networked systems. To address this limitation, \cite{Wu_Chen_Chou_Wang_2026} introduced the Fréchet Traffic Distance (FTD), a metric that compares real and synthetic network traffic in a multivariate feature space. By modeling both datasets as Gaussian distributions and computing the Fréchet distance between their statistical properties, FTD captures both distributional fidelity and temporal structure, making it particularly well-suited for applications such as network simulation and anomaly detection.

Building on this, more comprehensive evaluation frameworks adopt multi-dimensional approaches to assess synthetic data quality. \cite{Krishnan_Dhumpati_Salis_K_Sutaria_Abhyankar_2025} propose a combination of metrics—including Fréchet Inception Distance (FID), Wasserstein distance, feature consistency, pattern preservation, and precision/recall—to evaluate distributional, structural, and behavioral fidelity. Similarly, \cite{Ammara_Ding_Tutschku_2026} present a layered evaluation pipeline that separates fidelity, utility, and computational feasibility. Their approach first validates structural integrity using statistical comparison metrics before assessing downstream utility through intrusion detection performance, ensuring that high model accuracy does not obscure underlying data distortions. Collectively, these methods signal a move toward integrated evaluation frameworks that balance realism, practical utility, and scalability in synthetic cybersecurity data.

\subsection{Gaps}

Despite growing research in the fields of smart city data and SDG, several critical gaps continue to limit progress in smart city cybersecurity. First, most publicly available smart city datasets are operational rather than security‑focused, emphasizing domains such as transportation, energy, or environmental monitoring while offering little labeled attack data, network traces, or system‑level security events. Meanwhile, existing cybersecurity datasets typically originate from IoT or IIoT testbeds and fail to capture the scale, heterogeneity, and interconnectedness of real smart city environments. This disconnect makes it difficult to develop or validate cybersecurity tools that reflect actual smart city conditions.

Second, although SDG techniques such as GANs, statistical models, and machine‑learning‑based generators have advanced rapidly, their application to smart city contexts remains limited and fragmented. Current approaches tend to focus on isolated domains such as smart grids or enterprise networks rather than the multi‑domain complexity of smart cities. As a result, few frameworks can generate synthetic data that integrates multi-protocol and format cybersecurity data while preserving realistic device behavior, network interactions, and cross‑domain dependencies.

Finally, the SDG field lacks standardized benchmarks for evaluating outputs. Existing metrics capture aspects such as distributional similarity or task‑based utility but often overlook temporal dynamics, protocol compliance, and attack realism. There have been efforts to benchmark individual datasets, algorithms, or applications, but there are no universal thresholds for determining if a synthetic dataset is successful across the established evaluation factors, making it difficult to compare methods or validate results consistently.
\section{Methodology}

This section presents the research methodology used to design, develop, and evaluate the SDG (SDG) framework for smart city cybersecurity environments. Given the increasing complexity of smart city infrastructures and the challenges associated with collecting realistic, high-quality cybersecurity datasets, this research adopts a Design Science Research approach to guide the systematic creation and validation of an artifact capable of generating statistically valid and operationally useful synthetic datasets.

\subsection{Problem Identification}

The first phase focuses on identifying the core problem and establishing its significance within the domain of smart city cybersecurity. The first iteration of this phase has already been completed by reviewing existing cybersecurity datasets, analyzing their limitations, and highlighting the unique challenges of smart city environments that affect dataset availability such as data heterogeneity, interconnected systems, and privacy concerns. It was concluded that the absence of realistic and representative cybersecurity datasets limits the development, testing, and validation of smart city cybersecurity research and the development and testing of cybersecurity tools for smart city systems. It was also determined that existing, publicly-available smart city datasets primarily contain operational-type data and not cybersecurity data.

\subsection{Define Objectives of a Solution}

The overall goal is to close the gap in smart city cybersecurity dataset availability by using an AI-based framework to generate representative data. To achieve this, additional requirements will need to be set to determine the essential properties of a smart city cybersecurity dataset and to outline the desired properties of the AI-based SDG framework. Specific objectives include identifying the most relevant data types and features, determining minimum representational characteristics of smart city components, and establishing quality and validity criteria for the generated datasets. This phase will directly inform the design and development of the artifact.

\subsection{Design and Development}

This phase involves designing and building the primary research artifact: an AI-based framework capable of generating synthetic cybersecurity datasets for smart city environments. The framework integrates three key components: a data schema definition module (RQ1), a generative AI module for producing representative data (RQ2), and a malicious traffic generation module for injecting realistic malicious behavior (RQ3). The design will emphasize flexibility, reproducibility, and the ability to adapt to various smart city domains, ensuring that the artifact can generate data for multiple use cases and research needs.

\subsection{Demonstration}

In this phase, the developed artifact will be applied to demonstrate its utility and relevance. The SDG framework will be used to create sample smart city cybersecurity datasets that reflect real-world patterns and potential attack scenarios. The generated datasets will then be utilized to test existing cybersecurity tools, such as intrusion detection or anomaly detection systems, showcasing how the artifact can support both research and applied security testing. The demonstration will provide practical evidence that the artifact effectively addresses the identified problem.

Two demonstration scenarios are used to validate the artifact. The first scenario involves generating a benign synthetic smart city dataset for RQ2, demonstrating the model’s ability to reproduce normal traffic patterns. The second scenario addresses RQ3 by generating synthetic datasets with malicious activity embedded to simulate cyber attack conditions. These demonstrations evaluate the ability of the framework to generate context-aware synthetic data that supports both anomaly detection research and the testing of cybersecurity tools.

\subsection{Evaluation}

The evaluation phase assesses the performance, reliability, and validity of the developed artifact and its outputs. The generated datasets will be evaluated for statistical similarity to real-world data, diversity of simulated events, and the ability to support accurate cybersecurity tool performance. Quantitative evaluation methods will be used, including statistical analyses and benchmarking against existing datasets or detection models. The results will determine whether the artifact meets the research objectives and contributes meaningfully to the field.

\section{Proposed Framework}

\begin{figure}[ht]
\centering
\includegraphics[width=1\linewidth,keepaspectratio]{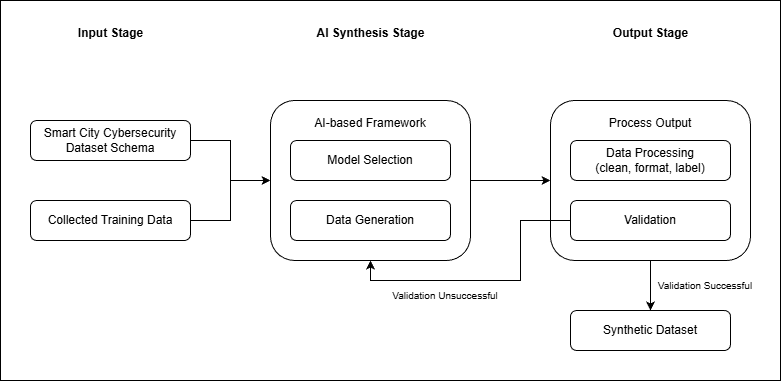}
\caption{Overview of the AI-based Synthetic Dataset Generation Framework}
\label{fig:SDG Framework}
\end{figure}

The research effort focuses on constructing an AI-based SDG framework capable of producing realistic smart city cybersecurity datasets. The framework is developed using an iterative, modular approach to ensure flexibility and continuous refinement. Each module is designed as an independent component that can be improved without modifying the entire system. The framework follows the DSR principles by ensuring the artifact is both rigorous in method and relevant to practical cybersecurity challenges.The framework’s operational workflow consists of three major stages: the Input Stage, AI Synthesis Stage, and Output Stage. Figure \ref{fig:SDG Framework} illustrates the high-level architecture of the framework, showing how input schemas and collected datasets are transformed into validated synthetic outputs.

\subsection{Input Stage} 
The Input Stage accepts a dataset schema definition model and collected IIoT datasets. This may be supplemented with data from smart city labs if available. The schema ensures structural consistency, while the real datasets provide contextual patterns and distributions to guide the AI-based synthesis.

The first task to be performed is performing a literature review to determine what constitutes a smart city cybersecurity dataset. Existing literature and datasets will be reviewed to determine what data types and values, features, protocols, and other characteristics should be present in such a dataset and hold to most impact for meaningful use. Existing, widely-used "traditional" cybersecurity datasets will also be reviewed for key features. This information will be used to create a smart city cybersecurity dataset schema (RQ1) which can be used as a blueprint for defining smart city cybersecurity datasets for this and other research. Concurrently, existing IIoT datasets will be collected to serve as the training data for the AI synthesis stage. 

With the schema developed and data collected, the next stage is data preparation. This includes labeling, cleaning, and feature extraction to convert raw packet captures or logs into structured formats suitable for training. Two types of datasets will be prepared: (1) benign datasets containing only normal smart city operational traffic and (2) malicious datasets that integrate cyber-attack scenarios such as scanning, spoofing, or denial-of-service behaviors. The malicious dataset is used to explore research question RQ3 by generating synthetic data that incorporates attack activity. 

\subsection{AI Synthesis Stage}

The AI Synthesis Stage is the core component of the SDG framework, responsible for transforming smart city cybersecurity training datasets into high-fidelity synthetic data. This stage leverages AI models that learn the statistical, structural, and behavioral patterns present in the input data and subsequently produce new data samples that adhere to the same constraints. The goal is to generate synthetic datasets that preserve the operational characteristics of smart city networks including device behaviors, communication patterns, and cyber-attack signatures while simultaneously protecting sensitive information and ensuring dataset realism.

\subsubsection{Model Selection}

The implementation begins with selecting generative model architectures suitable for time-series, event-driven, and heterogeneous smart city data. Candidate models include GANs (Packet GANs, Conditional Tabular GANs (CTGAN), etc.), Variational Autoencoders (VAEs), transformer-based time-series generators, and LLMs. Model selection will be driven by experimental comparison using baseline datasets and guided by the evaluation criteria defined in the evaluation plan.

\subsubsection{Schema and Constraint Integration}

To ensure domain validity, the AI models incorporate schema constraints, including field types (e.g., IP addresses, timestamps, sensor IDs), permissible ranges, and protocol-specific rules. These constraints are applied in two ways:

Pre-training encoding involves normalizing, embedding, or otherwise transforming structured fields into representations suitable for model input, while maintaining their semantic integrity.

Post-generation constraint enforcement ensures that synthetic outputs are validated and adjusted to comply with the operational rules of real-world smart city networks.

This dual approach ensures that the generated data reflects realistic system behavior while avoiding invalid outputs.

\subsubsection{Generation of Synthetic Data}

The AI models are trained using the preprocessed benign and malicious datasets produced in earlier stages. Once trained, the models will generate synthetic sample datasets with typical operational patterns such as periodic sensor updates, actuator commands, and standard network flows (RQ2). For RQ3, the models will also generate malicious data with attack scenarios such as scanning, lateral movement, spoofing, and other abnormal behavior intermixed with benign system activity.

\subsection{Output Stage}

The Output stage processes and validates the synthetic datasets, ensuring adherence to the input schema, labeling consistency, and statistical similarity with real-world distributions. Datasets are exported in standard formats suitable for intrusion detection, anomaly detection, and other cybersecurity research. The Output Stage acts as a gatekeeper, performing rigorous checks to confirm that the synthetic data adheres to schema requirements, reflects expected behavioral patterns, and maintains statistical fidelity to the underlying real-world smart city datasets.  

The process for generating smart city cybersecurity datasets --- both benign and those containing malicious or cyber-attack data (RQ3) --- will follow the same overall workflow. The primary difference lies in the input stage. For RQ2, the framework will use training data containing only benign activity collected from the smart city testbed to generate a normal dataset. In contrast, for RQ3, the input data will include samples captured during controlled attack scenarios within the testbed, allowing the framework to produce datasets with embedded malicious activity.

The framework will incorporate manual feedback mechanisms from validation metrics to iteratively improve the accuracy and utility of the generated datasets. Based on model performance in the AI Synthesis Stage and the validation results in the Output stage, the model selection stage will be re-executed with new models and/or parameters to achieve optimal results.

Once fully validated, the synthetic datasets are standardized and exported in formats commonly used across cybersecurity research and machine learning pipelines, such as CSV tabular flow-based data, PCAP for packet-level data requiring realistic payloads, JSON or NDJSON for IoT event logs and multi-modal sensor streams, and HDF5 for large-scale datasets that require efficient I/O access. Alongside the data, the export subsystem packages detailed metadata, including model and dataset versions, schema descriptions, evaluation metrics, generation parameters, timestamps, and experimental configurations. This provenance information ensures that the datasets remain fully reproducible, traceable, and consistent with best practices in scientific data management.

\section{Evaluation Plan}

Once the synthetic datasets are created, they will need to be evaluated on several dimensions to verify both their similarity to the real datasets and their usefulness for cybersecurity tools and testing. The evaluation strategy integrates statistical similarity testing, fidelity and realism analysis, and task-based performance evaluation to determine the utility of the generated datasets for intrusion detection and related cybersecurity applications. 

\subsection{Schema Adherence and Structural Validation}

The first component of the Output Stage is dedicated to ensuring that all synthetic records align with the schema of the original dataset. This process involves several layers of validation: required fields must be present and correctly typed, such as timestamps, protocol identifiers, MAC/IP addresses, and sensor values; each field must adhere to defined ranges or categorical sets, including protocol numbers, sensor thresholds, and device IDs; formats are checked for compliance with standards like IP and MAC address structures, time window constraints, and protocol-specific rules; and cross-field consistency is verified to confirm that protocol combinations and flow-direction indicators logically correspond to expected device roles. Records that fail these schema checks are either repaired using rule-based heuristics or excluded altogether to maintain the integrity and quality of the final dataset. If significant failures are present, the Input and/or AI Synthesis Stages will be revisited and parameters updated.

\subsection{Statistical Similarity}

The next phase is to verify statistical similarity of the synthetic datasets to the real data. The synthetic data will be validated for statistical similarity, fidelity and realism. This will be performed with existing open-source tools such as SDNist, SDMetrics, SynthEval, and SynthRO. None of these tools were specifically designed for IoT or smart city data, so each will need to be evaluated to determine the best tool and metrics within the tool for the smart city use case.

\subsection{Labeling and Annotation Consistency}

For datasets supporting intrusion detection and cyber-attack analysis, preserving label integrity is crucial. In the Output Stage, automated checks verify that benign and malicious labels are correctly assigned based on synthesis inputs, that attack type labels remain consistent with intended categories such as DoS, spoofing, or scanning, and that multi-step attack sequences carry coherent labels across consecutive packets, flows, or events. These safeguards prevent mislabeled records that could compromise training pipelines or reduce the reliability of IDS evaluations.

\subsection{Utility Testing}

If the synthetic data passes the previous evaluation checks, the next phase is to determine utility of the data by attempting to use the datasets for real-world cybersecurity tasks. For RQ3, creating the ability to generate malicious data, this will be essential to validate that malicious data was successfully generated. The datasets will be fed into open-source security tools such as intrusion detection systems to determine if the data generation was successful. ICS/SCADA specific security tools, such as Idaho National Lab's (INL) Malcolm network analysis tool \cite{idaholab/Malcolm_2026} will be utilized for their specialized support for IIoT protocols. 

\subsection{Combined Benchmark}

One outcome of this research will be the development of a comprehensive benchmark that integrates schema adherence and structural validation, statistical similarity, labeling and annotation consistency, and utility testing into a single meaningful metric. This benchmark will be used to assess whether the framework generates representative synthetic data. The Fréchet Traffic Distance (FTD), Fréchet Inception Distance (FID), and Wasserstein distance techniques discussed earlier will be evaluated using synthetic smart city datasets to determine their applicability to this domain. If these techniques prove insufficient, modifications or new benchmarking methods will be proposed to more accurately measure the representativeness of synthetic smart city data.

\section{Expected Contributions}

This research is expected to advance smart city cybersecurity by providing a repeatable framework to create high‑fidelity synthetic datasets that enable rigorous testing, training, and evaluation of security tools. The framework will provide a scalable foundation for developing and validating intrusion detection systems, anomaly‑detection models, and other cybersecurity analytics tailored towards smart city applications. City technology teams, vendors, and IT staff will also benefit from synthetic datasets that support system testing, incident‑response exercises, and team training in safe, controlled environments, ultimately improving the resilience of critical smart city services.

The research is also expected to contribute to the ongoing development of evaluation methodologies for synthetic data. Although several metrics have been proposed for assessing synthetic data quality, including techniques designed for network traffic datasets, there remains limited guidance regarding what constitutes an acceptable or representative score for these measures. While this work is not intended to establish definitive benchmarks or standards, it will provide empirical results that can inform future efforts to develop meaningful evaluation criteria for synthetic smart city data.

Additionally, the framework will be validated using a representative subset of smart city protocols and applications to demonstrate its effectiveness in modeling the dynamic and heterogeneous characteristics of smart city environments. While it is not feasible to evaluate every smart city technology within the scope of this research, the proposed methodology is designed to be adaptable to additional protocols, devices, and application domains. By demonstrating the framework across several representative data sources, this research will establish a foundation for extending synthetic data generation techniques to other smart city systems with similar complexity and variability.

Finally, beyond immediate technical impacts, the research contributes to workforce development and cross‑sector collaboration. The synthetic datasets can be used to build hands‑on labs and training modules for students, researchers, and practitioners, helping prepare the next generation of cybersecurity professionals to address the unique challenges of smart city systems. Because generated datasets will be fully synthetic and free of sensitive information, they can be shared widely without risking citizen privacy or exposing critical infrastructure details, enabling broader collaboration across academia, industry, and government. Collectively, these contributions strengthen the safety, resilience, and research readiness of smart cities.

\section{Current Status}

This research is currently in the data collection and exploratory analysis phase, with a focus on identifying and evaluating relevant IIoT cybersecurity datasets to serve as training data to the SDG framework. Initial efforts involve performing exploratory data analysis to understand dataset structure, feature distributions, protocol representations, and the presence of malicious activity. In parallel, a lab environment is being developed to support both data generation and evaluation activities. This includes provisioning a dedicated virtual machine configured to run INL's Malcolm for network traffic analysis and utility testing. Additionally, work is underway to integrate this environment with a GPU-enabled compute cluster managed with Slurm to enable experimentation with computationally intensive SDG models. Feedback is particularly sought on the dataset schema definition, the choice of generative modeling approaches, and the proposed criteria for assessing realism and downstream cybersecurity utility.

\section{Scope}

This study focuses on the design, development, and evaluation of an AI-driven SDG framework for smart city cybersecurity research. The study is limited to modeling cyber-physical data relevant to smart city infrastructures, which may include network traffic, IoT device telemetry, system logs, and simulated adversarial activity.

The research examines whether synthetic data can achieve sufficient structural, statistical, and behavioral fidelity to support cybersecurity tool training and evaluation. Evaluation is conducted through quantitative performance metrics, statistical similarity analysis, and adversarial scenario testing.

The study does not attempt to replicate an operational smart city environment. Instead, it focuses on subsystems and data commonly found in smart city infrastructures, such as interconnected IoT networks and associated cybersecurity event logs.

The primary outcome of the study is the validated SDG framework and the evaluation results demonstrating its utility for cybersecurity research and testing purposes.

\section{Limitations}

Several limitations may affect the findings of this study. First, the availability and quality of existing smart city and cybersecurity datasets may impact the identification of the defining characteristics necessary for synthetic replication. If source datasets are incomplete or biased, the synthetic data may inherit those limitations. Second, synthetic data may not capture the full complexity of real-world smart city environments. As a result, tool performance observed in synthetic environments may not fully generalize to live operational systems. Third, AI-based generative models may introduce unintended artifacts, statistical distortions, or oversimplifications that influence evaluation outcomes. Detecting and mitigating such distortions may not be fully achievable. Finally, computational and storage resource limitations may restrict the scale or complexity of generated datasets, potentially affecting realism in large-scale infrastructure simulations.

\section{Future Work}

Future work will focus on the implementation and evaluation of the SDG framework tailored to smart city cybersecurity environments. Building on the insights gained from exploratory data analysis, the next phase will involve defining the most meaningful properties of a smart city cybersecurity dataset, including relevant features, protocols, and attack scenarios. Subsequently, multiple SDG techniques will be implemented and evaluated to determine their effectiveness in capturing the complex nature of smart city systems.

In parallel, efforts will be directed toward integrating realistic attack simulation into the SDG process to ensure the inclusion of meaningful malicious behavior for cybersecurity tool development. Additionally, a comprehensive evaluation framework will be established, incorporating metrics for statistical fidelity, temporal and behavioral realism, and cybersecurity tool utility in security applications such as intrusion detection and anomaly detection.

\section{Conclusion}

Smart city infrastructures pose substantial cybersecurity challenges due to their scale, diversity, and dependence on interconnected cyber‑physical systems. Yet progress in securing these environments remains limited by the scarcity of high‑quality, representative datasets. Existing datasets are either operational in nature and lack security relevance or are derived from isolated IoT/IIoT environments that do not fully capture the complexity of smart city ecosystems.

This research addresses these challenges by introducing an AI-driven SDG framework tailored to smart city cybersecurity needs. By leveraging SDG techniques to create realistic, high-fidelity datasets, this work aims to close the gap between data scarcity and the need for smart city cybersecurity research and tool development. In addition to generating synthetic data, this research emphasizes the importance of incorporating realistic attack scenarios and establishing rigorous evaluation methodologies to ensure that the generated datasets are both reliable and useful for downstream applications.

The contributions of this study include defining the essential characteristics of smart city cybersecurity datasets, applying and assessing SDG techniques in complex, multi‑domain environments, and developing comprehensive strategies for evaluating synthetic data quality. Collectively, these efforts establish a foundation for reproducible smart city cybersecurity research and support the broader goal of strengthening the security and resilience of smart city infrastructures by equipping researchers and practitioners with realistic, shareable data for understanding emerging threats.

\printbibliography

\end{document}